\documentclass[twocolumn,showpacs,preprintnumbers,amsmath,amssymb]{revtex4}
\usepackage{tabularx,graphicx}\begin{document}
\newcommand{\beq}{\begin{equation}}
\newcommand{\eeq}{\end{equation}}
\newcommand{\beqn}{\begin{eqnarray}}
\newcommand{\eeqn}{\end{eqnarray}}
\newcommand{\bmath}{\begin{subequations}}
\newcommand{\emath}{\end{subequations}}
\title{Did Herbert Fr\"{o}hlich predict or postdict the isotope effect in superconductors?}
\author{J. E. Hirsch }
\address{Department of Physics, University of California, San Diego\\
La Jolla, CA 92093-0319}
 
\begin{abstract} 
Herbert Fr\"{o}hlich is generally credited with having predicted   the  fundamental role of  electron-phonon interactions   in superconductivity in 1950, and in particular the isotope effect,
without any experimental input. Here we examine the facts on which this belief is based and point out that whether or not 
the generally accepted view is true depends on the meaning of the word $shortly$.
 \end{abstract}
\maketitle 

\section{the sequence of events}

On May 16th, 1950, Herbert Fr\"{o}hlich's paper entitled  ``Theory of the Superconducting State. I. The Ground State at the Absolute
Zero of Temperature'' was received by the Physical Review\cite{fr1}.  The paper (hereafter called `May 16th paper') proposed that the interaction between electrons and lattice
vibrations  (phonons) was
responsible for superconductivity.  
The paper made no mention of the fact that in the May issue of the same journal two experimental 
papers\cite{reynolds,maxwell} (both received by the journal on March 24th, 1950) were describing
measurements of  the critical temperature of mercury for different isotopes, reporting  that
{\it ``there is a systematic decrease of transition temperature with 
increasing mass''}\cite{reynolds} and that\cite{maxwell} {\it ``From these results one may infer that the transition temperature
of a superconductor is a function of the nuclear mass, the
lighter the mass the higher the transition temperature. ''}

Three days later, on May 19th, 1950, a  ``Letter to the Editor''
 by Fr\"{o}hlich entitled ``Isotope Effect in Superconductivity''  was received by the journal Proceedings of the Physical Society of
London\cite{fr2}. In that short note
(half a page long), Fr\"{o}hlich stated that the isotope effect experiments {\it ``have just come to my notice''}
 and pointed out that the formalism in his May 16th paper in fact predicted the effect. 
 In a note added in proof to the May 16th paper
Fr\"{o}hlich states {\it  `` The isotope effect [see Reynolds et al. , Phys. Rev. {\bf 78}, 487 (1950); 
Maxwell, Phys. Rev. {\bf 78}, 477 (1950)] which has recently come to my notice follows quantitatively from the proportionality of [S] with the inverse isotopic mass  $1/M$ [ see e.g. Eq. (6.9) where F depends on $Ms^2$  only and is hence independent of the isotopic mass]  as was stated in a recent note [Fr\"{o}hlich, Proc. Phys. Soc. {\bf A63}, 778 (1950)]. 
 This agreement provides a direct check for the fundamental assumptions of the
theory. ''} 

A competing explanation of the isotope effect experiments was proposed by John Bardeen in a short paper received by 
Physical Review just three  days later, 
 on May 22nd, 1950\cite{bardeen1}. Bardeen had no knowledge of Fr\"{o}hlich's work at that time, but had just been made
aware of the isotope effect experiments in a telephone call by B. Serin on May 15th\cite{frseminar}. In his paper Bardeen readily acknowledged this,
with the words {\it ``This approach was stimulated
by the recent finding of Reynolds, Serin, Wright, and Nesbitt
and of E. Maxwell that the critical temperature, $T_c$, of mercury depends on the isotopic mass.''}

About a week after   Bardeen had submitted his paper,
Fr\"{o}hlich visited Bell Laboratories (where Bardeen was at the time)
  where he discussed  his work on the electron-phonon interaction and
the isotope effect\cite{frseminar}. This is when Bardeen learned for the first time about Fr\"{o}hlich's work on the
subject. 
According to Bardeen's recollection\cite{bardeenfrseminar},  Fr\"{o}hlich stated then that he had developed his theory
{\it `` without knowledge of the isotope effect''}. To this author's knowledge, this occasion 
was the first recorded time when  Fr\"{o}hlich's work on the subject became known to anybody other
than Fr\"{o}hlich himself  
 and the journals where Fr\"{o}hlich's had submitted his papers  a few days earlier.

Starting with these events\cite{crystal}, the electron-phonon interaction became the focus of theoretical efforts attempting to explain
superconductivity after 1950, culminating in the BCS theory in 1957, generally believed to explain `conventional' superconductivity. 
Neither Fr\"{o}hlich's nor Bardeen's 1950 theories stood the test of time. Nevertheless, based on the events of May 1950 recounted
above,  Fr\"{o}hlich is generally credited with having  $predicted$ the fundamental role of electron-phonon interactions in 
superconductivity and the isotope effect without any experimental input. In this paper we analyze the evidence in support of this interpretation of events
and raise the possibility that there could be alternatives.

\section{Fr\"{o}hlich's version of events}

In numerous papers following the ones cited above, Fr\"{o}hlich states that he developed his theory independent
 of the isotope effect
experiments. For example:

\noindent 1951\cite{fr1951}: {\it ``The prediction that it is the interaction between electrons and lattice vibrations$^1$ which is responsible for the
establishment of the superconducting state has been verified by the discovery of the isotope effect$^2$.''}(Ref. 1 is
Fr\"{o}hlich's May 16th paper.)

\noindent 1952\cite{fr1952}: {\it ``...the situation is best described in terms of a field theory in which the
electrons are the sources of the vibrational field. Discussion with the help of perturbation theory led to the 
introduction of an interaction parameter $F$. It was found that if $F$ is larger than a critical value
$F_0$ then the electron distribution in momentum space differs in the ground state from the
normal distribution. This new state was tentatively indentified with the superconductive state which
led to a prediction of the isotope effect. Starting from a knowledge of this effect, Bardeen (1950) has 
developed a theory on similar lines.''}

\noindent (Note how Fr\"{o}hlich makes a clear distinction between Bardeen's work that was started with
knowledge of the isotope effect, and his own.)

\noindent 1953\cite{fr1953}: {\it ``expression (1.5) predicts that the energy difference S should
vary as 1 /M. This prediction was confirmed quantitatively with the
discovery of the isotope effect...It seems, therefore, that the isotope effect represents a
most direct confirmation for the hypothesis advanced above.''}

\noindent 1954\cite{fr1954}: {\it ``The conjecture that the interaction between electrons coupled through the field of lattice displacements is responsible for superconductivity (Frohlich 1950) has been strongly supported by the discovery of the isotope effect (Maxwell 1950; Reynolds, Serin, Wright  and Nesbitt 1950; Bar, Mendelssohn, Olsen, Allen and  Dawton 1950; Lock, Pippard, Shoenberg, Allen and Dawton 1950).''}

\noindent 1954\cite{fr1954b}:  {\it ``In fact after the
first, application of the methods of field theory to electrons in ionic crystals
(Fr\"{o}hlich, Pelzer and Zienau 1950), the use of these methods in metals
led to an important step in the theory of superconductivity and to the
prediction of the isotope effect (Fr\"{o}hlich 1950).''}

\noindent 1961\cite{fr1961}:  {\it ``While thus a satisfactory solution was long delayed, a most important feature
of the interaction was already discovered at the first attempt. It led to the
prediction of the isotope effect (Fr\"{o}hlich 1950) which was discovered soon afterwards
(Maxwell 1950, Reynolds et al. 1950, Allen et al. 1950 a, b).''}

\noindent 1963\cite{fr1963}: 
\noindent  {\it ``The recognition by the author that the electron-phonon
interaction in metals responsible for normal resisitivity also induces an electron-electron interaction, in conjunction with the suggestion that this interaction be responsible for superconductivity, transformed the till then exasperatingly vague task of developing the theory of superconductivity into a definite mathematical problem expressed in terms
of a characteristic Hamiltonian. Discovery of the
isotope effect confirmed this suggestion and, on
the basis of a simple model, the mathematical
problem found a successful solution in the BCS
theory.''}

Furthermore, in the 1961 review paper\cite{fr1961} Fr\"{o}hlich stressed that his proposal was completely independent of any other ideas being considered at the 
time: 
\noindent  {\it ``A dictum was
accepted nearly universally, namely that the ions of the lattice, in view of their
large mass (compared with electrons), could play no important role in the establishment
of the superconductive state. It was in the face of this opinion that the
author conceived the idea (Fr\"{o}hlich 1950) that just the opposite of the ÔdictumÕ
contains the truth.''}
\noindent And later in this paper he states: {\it ``, the development of the
theory of superconductivity is an excellent example of the intuitive approach.
My idea was foremost that one should consider the small energy involved in the
superconductive transition as the first problem to be answered. Electron-phonon
interaction provided this possibility from a purely dimensional point of view.}

\section{The community's version of events}
There appears to be universal agreement in  the community that Fr\"{o}hlich's theory was developed independent
of the experiments. This author could not find a single instance in the literature suggesting otherwise.

For example, this version of events was rapidly adopted  by the authors of the isotope effect experiments:

\noindent  E. Maxwell, 1952\cite{maxwell52}: {\it ``Furthermore, theoretical treatments by Fr\"{o}hlich$^{13}$
 and Bardeen$^{14}$
appeared, which were based on interaction
between electrons and lattice vibrations and which
specified that $T_c \sim M^{-1/2}$. (Fr\"{o}hlich had developed his
theory prior to his knowledge of the isotope effect. )''}

\noindent B. Serin, 1955\cite{serin55}: {\it ``In fact Fr\"{o}hlich$^{26}$, before hearing of the experimental
investigations, had suggested that superconductivity comes about as the result of just such an
interaction. (Conversely, the experimentalists were unaware of Fr\"{o}hlich's theory.)''}

Superconductivity textbooks uniformly credit  Fr\"{o}hlich with predicting the isotope effect without experimental
input. For example:

\noindent  Blatt, 1964\cite{blatt}: {\it ``Whereas the distribution of superconductors in the periodic table was
known, to some extent, before the work of Fr\"{o}hlich, the isotope effect was not known to Fr\"{o}hlich and thus
constituted a sigmificant theoretical prediction.''}

\noindent  Schrieffer, 1964\cite{schrieffer}: {\it ``It was not until 1950 that the basic forces responsible for the
condensation were recognized, through the insight of Fr\"{o}hlich.$^{10}$ He suggested that an
effective interaction between electrons arising from their interaction with crystal lattice vibrations (phonons)
was of primary importance in bringing about the condensation. At this time, independent experiments on the
isotope effect in superconductors were being carried out by Reynolds et al$^{11}$ and by Maxwell$^{12}$
which gave experimental support to Fr\"{o}hlich's point of view.''}

\noindent  Rickayzen, 1965\cite{rickayzen}: {\it ``... On the basis of this idea, he predicted the isotope effect independently
at the same time as it was being discovered experimentally.''}

\noindent  Tinkham, 1975, 1996\cite{tinkham}: {\it ``Historically, the importance of the electron-lattice interaction in
explaining superconductivity was first suggested by Fr\"{o}hlich$^4$ in 1950.  This suggestion was confirmed
experimentally by the discovery$^5$ of the isotope effect, i.e. the proportionality of $T_c$ and $H_c$ to $M^{-1/2}$ for 
isotopes of the same element.''}

In the appendix, we give several more examples of statements by prominent experimentalists and theorists in the 
field over the years expressing the same view on the issue.

L. Hoddeson, a historian of physics who has written extensively on the early history of superconductivity and
on  John Bardeen, writes\cite{hoddeson}:
{\it ``As it happened, Bardeen was not the only theorist to connect superconductivity with the electron$-$lattice interaction. Earlier in 1950, before
Maxwell and Serin found the isotope effect experimentally, Herbert
Fr\"{o}hlich had set forth a theory predicting it. When Fr\"{o}hlich learned of the
experimental results a day or two after they appeared in the Physical
Review, he sent a letter to the Proceedings of the Royal Society to claim
priority for his theory. The competition was on.''}

However, upon inquiring with L. Hoddeson on the meaning of the
 ``set forth'' part of the statement above, Hoddeson could not identify a source for it and
 acknowledged that it is possible the 
statement may have been  an interpretation of statements by others rather than based on direct
evidence (private communication with this author).

\section{Bardeen' s early version of events}

John Bardeen was consistent in giving Fr\"{o}hlich full credit for the independent prediction of the isotope effect,
both before and after the BCS work.   According to the ISI Web of Science database, Bardeen cited Fr\"{o}hlich's May 16th paper on
13 different occasions. On the other hand he never cited  Fr\"{o}hlich's May 19th note\cite{fr2} that explicitly
addressed the isotope effect, presumably because in Bardeen's view it was unnecessary, the prediction being clearly implicit in
the May 16th paper. The following are examples of how Bardeen referred to  Fr\"{o}hlich's work:

\noindent 1951\cite{bardeen1951}    {\it ``Prior to his knowledge of the isotope effect,
Fr\"{o}hlich developed a theory of superconductivity
based on the self-energy of the electrons arising from
interactions with the phonon field.''}

\noindent 1956\cite{bardeen1955}    {\it ``Fr\"{o}hlich's theory, developed without knowledge of the isotope
effect, gave a relation between critical temperature, $T_c$, and isotopic mass
$\sqrt{M} T_c\sim const$ which is close to that found empirically.''};  later in the same article: {\it ``Without having prior knowledge of the isotope effect, Fr\"{o}hlich [4] proposed a theory of superconductivity
based on electron-phonon interactions.''}

\noindent 1957 (BCS)\cite{bcs}:
{\it ``A great
breakthrough occurred with the discovery of the isotope
effect,$^{10}$ which strongly indicated, as had been
suggested independently by Fr\"{o}hlich,$^{11}$ that electron-phonon
interactions are primarily responsible for
superconductivity.''}

\noindent 1962\cite{bardeen1962}    {\it ``As you know, the basis for
the current theory was suggested by Fr\"{o}hlich in 1950
and confirmed by the simultaneous discovery of the  isotope
effect. ''}

\noindent 1969\cite{bardeen1969}
{\it ``In 1950 Herbert Fr\"{o}hlich$^8$ made a proposal that led eventually to the pairing theory of
superconductivity. He based his theory on interactions between electrons and phonons, the quanta of the
lattice vibrations. That year also two groups independently discovered that the superconducting transition
temperature depends on isotopic mass.''}

\section{Bardeen's late version of events}

In the introduction to his 1972 Nobel Prize lecture\cite{bardeennobel}  Bardeen wrote, 
consistent with his earlier statements on the subject:

\noindent {\it ``The year 1950 was notable in several respects for superconductivity theory.
The experimental discovery of the isotope effect [4, 5] and the independent
prediction of H. Fr\"{o}hlich [6] that superconductivity arises from interaction
between the electrons and phonons (the quanta of the lattice vibrations) gave
the first clear indication of the directions along which a microscopic theory
might be sought.''}

\noindent  However, in the next section he wrote:

\noindent  {\it ``The isotope effect was discovered in the spring of 1950 by Reynolds, Serin,
et al, [4] at Rutgers University and by E. Maxwell [5] at the U. S. National
Bureau of Standards. Both groups measured the transition temperatures of
separated mercury isotopes and found a positive result that could be interpreted
as $T_cM^{1/2}\sim$  constant, where $M$ is the isotopic mass. If the mass of the ions
is important, their motion and thus the lattice vibrations must be involved.
Independently, Fr\"{o}hlich, [6] who was then spending the spring term at
Purdue University, attempted to develop a theory of superconductivity based
on the self-energy of the electrons in the field of phonons. He heard about
the isotope effect in mid-May, shortly before he submitted his paper for
publication and was delighted to find very strong experimental confirmation
of his ideas.''}

  This is a very significant departure from the previous version of events. For the first time, it is stated
that Fr\"{o}hlich knew about the isotope effect experiments $before$ he submitted his May 16th paper!

This new version of events is further supported by the paper written by Bardeen for Fr\"{o}hlich's
$Festschrift$ on occasion of his retirement\cite{festschrift} in 1973. On one hand, Bardeen wrote
\noindent {\it `` Fr\"{o}hlich did his original work on electron-phonon interactions and superconductivity while he
was visiting at Purdue University in the spring semester of 1950. Independently, without his
knowledge, two groups had been measuring the transition temperatures of separated mercury
isotopes and found a positive result that could be interpreted as $T_c M^{1/2}=$const, where $M$ is
the isotopic mass.''}  In the next paragraph, Bardeen wrote 
that the experimental 
papers reporting the isotope effect {\it ``appeared in mid-May. It was not until this time, just before he was ready to submit his own paper for publication, that Fr\"{o}hlich learned about this strong experimental confirmation of his ideas.''}

Thus, Bardeen stated twice in the literature unequivocally that Fr\"{o}hlich knew about the isotope effect experiments
before he submitted his May 16th paper {\it that did not mention the isotope effect experiments}.

In the same paper\cite{festschrift}, Bardeen recounts that the experimental results on the isotope effect had been reported at a
small meeting almost two months earlier:

{\it ``At that time, a great deal of research in low temperature physics in the United States was sponsored by the US Office of Naval Research, and periodic meetings were held to exchange information. The isotope effect was first announced by the two groups at an ONR sponsored meeting held at the Georgia Institute of Technology in March, 1950, but was not publicized outside of the low temperature community. Papers were submitted for publication shortly after the meeting, and appeared in mid-May.''}

According to a report on that meeting by W. T. Ziegler published in Science magazine in May 1950\cite{scienceonr}, the meeting was held March 20$-$21
and attended by more than 60 scientists from 24 institutions. On this topic, the Ziegler report states {\it ``Among the interesting
experimental results reported in the field of superconductivity were those on the mercury isotopes. 
Experiments on these isotopes indicated that the normal transition temperature decreased by
$0.01^o$ per unit increase in mass number.''}

\section{Fr\"{o}hlich's own recount of events}
In 1982, Fr\"{o}hlich wrote in the proceedings of a NATO Advanced Study Institute on Advances in Superconductivity\cite{advances}:

\noindent
{\it `` The isotope effect in superconductors was in fact
discovered during the same period. At that time I was at
a 2-3 month visit to Purdue University, and submitted my
paper on leaving on the 16.5.1950. I then spent a couple
of days at Princeton and there, at my breakfast table,
found the Physical Review with the two letters reporting
the isotope effect [12], [13]. On checking I found my
M-dependence confirmed and on the 19.5.1950 sent a letter
[ 14 ] to claim confirmation of the basic idea, electron-
phonon interaction. ''}

Clearly, the statement is intended to imply that Fr\"{o}hlich did not know about the isotope effect experiments
before he read the Physical Review at his breakfast table in Princeton. 
However  note that it does not say so 
explicitly. If it did, it would be in direct contradiction to the statements of John Bardeen 
recounted in the previous section, since Fr\"{o}hlich went to Princeton right $after$ submitting his
May 16th paper. (This statement also suggests there  may be some confusion in Fr\"{o}hlich's mind between
submission and receipt dates, but this is irrelevant to the issue at hand).  

To this  author's knowledge, this statement is the only one ever made by   Fr\"{o}hlich in the literature on how he learned about the isotope effect
experiments. The statement leaves us completely in the dark as to whether, and if yes 
where, when and how he may have  {\it ``heard about the isotope effect in mid-May, shortly before he submitted his
paper for publication''}, as affirmed by Bardeen in his   Nobel lecture.

\section{Discussion  of the facts}

Note that the fact that Fr\"{o}hlich was visiting Purdue University while developing his theory of superconductivity, as opposed to
being in his home base in Liverpool, makes absolutely no difference within the generally accepted version of events. 
Instead, if Fr\"{o}hlich learned about the isotope effect experiments through an informal channel rather than by
reading the Physical Review, this would have been presumably  facilitated by being in the U.S. rather than in England
during that time.

This author sees no reason to doubt that Bardeen's statements that Fr\"{o}hlich learned about the isotope effect experiments
$before$ he submitted his May 16th paper for publication were factually accurate. Recall that one of these statements was in Bardeen's 1972 
Nobel lecture,
an unlikely place to make such a remark lightly. If those statements had been factually inaccurate, it is
reasonable to expect that Fr\"{o}hlich would have explicitly challenged them. No such challenge exists in the
literature to this author's knowledge.  Fr\"{o}hlich passed away in 1991, 18 years after Bardeen put forth this version of
events, and it is inconceivable  that Fr\"{o}hlich would not have been aware of these statements,
one of them made by Bardeen on Fr\"{o}hlich's own
$Festschrift$. 

How Bardeen found out that Fr\"{o}hlich knew about the experiments before submitting his paper, and why Bardeen chose
to disclose this information in his 1972 and 1973 papers, are interesting questions that probably have
interesting answers.

Assuming then that  indeed Fr\"{o}hlich was aware of the experiments 
{\it `` just before he was ready to submit his own paper for publication''},
 it is rather
peculiar that he made no mention of them in the paper,
nor of the fact that he was {\it ``delighted to find very strong experimental confirmation
of his ideas''}, as stated by Bardeen.

There is no source given in Bardeen's statement to his assertion that Fr\"{o}hlich learned about the experiments
only $shortly$ before submitting his May 16th paper.
The fact that a meeting was held two months earlier attended by over 60 scientists where the experimental
results were reported makes it
at least plausible that Fr\"{o}hlich could have heard about the experiments quite a bit earlier than ``shortly'' before
submitting his paper.  Note that Bardeen produced
a theory proposing an explanation for the experiments in a period of less than 7 days following the telephone call from Serin 
(May 15th was the telephone call, May 22nd the receipt date of Bardeen's paper by the journal.)

We conclude that  Bardeen's source of information that Fr\"{o}hlich learned about the isotope effect experiments
only $shortly$ before submitting his paper, after Fr\"{o}hlich's theory had been completed, is likely to have been
 statements by  Fr\"{o}hlich himself to Bardeen. 
If Fr\"{o}hlich was
truthful, indeed he predicted the isotope effect and the fundamental role of electron-phonon interactions 
in superconductivity independent of the  isotope effect
experiments. If Fr\"{o}hlich was not truthful, he did not.

If Fr\"{o}hlich was truthful, the question remains: why didn't he mention the isotope effect experiments in any
way, shape or form, in his May 16th paper?

If Fr\"{o}hlich was not truthful, many questions remain: what did Fr\"{o}hlich know, and when did he know it? From what source(s)?
How did this knowledge influence his work? And how (if at all) did the distortion of the real facts influence the development of the field?

From a purely probabilistic point of view, the probability that the two events,  Fr\"{o}hlich submitting
his May 16th paper, and Fr\"{o}hlich learning about the isotope  effect experiments, would happen within 
a short time interval of each other, $if$ the two events are truly independent of each other, becomes increasingly unlikely the shorter the time interval
between the two  events
relative to the total time from the discovery of superconductivity on April 8th, 1911
to May 16th, 1950, namely  14,283 days.

This author believes that the true version of events is likely to have been {\it very} different from the generally accepted
version. In a subsequent publication I will discuss an alternative version and arguments in support of it, as well
as argue that the fact that the accepted version may not have corresponded to the true
facts had a very substantial  detrimental effect on the course of   development of  the
field.

\acknowledgements  The author is grateful to F. Marsiglio for a discussion that stimulated his interest in this
subject,   to L.J. Sham and S.K. Sinha for discussions and feedback, and to L. Hoddeson for a 
private communication.

\section{appendix}
The following are some examples of statements by prominent experimentalists and theorists in the 
field   over the years expressing the view that Fr\"{o}hlich's work was completely independent of the isotope effect 
experiments. We have not found a single instance in the literature with a differing viewpoint.

\noindent Tisza, 1950\cite{tisza}: {\it `` The present manuscript was essentially completed
when Fr\"{o}hlich's$^{23}$  theory of superconductivity came to
the author's attention. (Compare also the recent note of
Bardeen.$^{24}$) This theory is based on the interaction of
free electrons with the lattice vibrations and led to a
brilliant prediction of the isotope effect.''}

\noindent Peierls, 1959\cite{peierls}: {\it ``In 1950, Fr\"{o}hlich suggested that superconductivity might be due
to an interaction between different electrons in the metal which was transmitted through the vibrations of the
crystal lattice. ... Fr\"{o}hlich's view was strikingly confirmed by the discovery about the same time of the 
isotope effect''}.

\noindent Ginsberg, 1962\cite{ginsberg62}: {\it ``It is interesting to note that slightly prior to the publication of the
experimental indications of the isotope effect, it had been suggested that the electron-phonon
interaction might be responsible for superconductivity$^{20}$.''} (Ref. 20 is to Fr\"{o}hlich's May 16th paper).

\noindent   Matthias, 1964\cite{matthias}: {\it  ``Fr\"{o}hlich predicted a gap in the electron spectrum on the top of
the Fermi surface in the superconducting state for a one-dimensional model. This gap was indeed detected
optically, by microwave spectroscopy (6) as well as by specific heat measurements (7). He also
predicted the isotope effect, according to which the transition temperature $T_c$ varies with the
inverse square root of the atomic mass, thus clearly indicating a correlation with lattice vibrations.''}

\noindent Cleason and Lundqvist, 1974\cite{cleason}: {\it '' An important step 
towards a microscopic theory was taken by the theoretical prediction of the isotope effect by Fr\"{o}hlich [11] and 
the independent experimental discovery of Maxwell  and 
Reynolds et al. [12, 13].''}

\noindent Pippard, 1987\cite{pippard} {\it ``It was a notable triumph for Fr\"{o}hlich that he predicted the isotope
effect which was so speedily confirmed by four different groups$^{72}$.''}

\noindent Mott, 1991\cite{mott}: {\it ``As the phonon frequency enters into the expression, Fr\"{o}hlich's theory showed that 
the transition temperature should depend on the isotopic mass of the metal concerned, as was found to
be valid experimentally by several workers at about the same time. Fr\"{o}hlich immediately saw that these results
were predicted by his work.''}

\noindent Schrieffer, 1992\cite{schrieffer92}: {\it ``Independently and without knowledge of the experimental
results on the isotope effect, Herbert Fr\"{o}hlich attempted to deduce the Meissner effect by proposing a 
theory of superconductivity based on the electron-phonon interaction.$^2$''}

\noindent Geballe, 1993\cite{geballe}: {\it ``The discovery of the inverse square-root dependence of
$T_c$ upon mass for the different isotopes of tin, lead and mercury in 1950, by four different groups in the US
and Britain, restimulated Bardeen's intense interest in the electron-phonon interaction. Evidence for this
isotope effect, predicted in a model proposed in that same year by Herbert Fr\"{o}hlich,$^6$ had actually been
sought experimentally at Leiden in the 1920s, when Pb isotopes first became available.''}

\noindent Schrieffer and Tinkham, 1999\cite{st}: {\it 
``The discovery of the isotope effect by Maxwell (1950) and Reynolds et al (1950), namely that
$T_c\sim M^{-\alpha}$ where $M$ is the ionic mass and $\alpha \sim 1/2$, gave strong support to the
view that superconductivity is the result of the electron-phonon interaction. Prior to this discovery,
Fr\"{o}hlich (1950) had worked out a model based on this interaction but ran into formal difficulties and the
approach did not describe the properties of a superconductor.''}

\noindent Goodstein and Goodstein, 2000\cite{goodstein}: {\it ``
The foundations that would eventually lead to a microscopic theory began to be
laid down, also around 1950, when Herbert Fr\"{o}hlich realized that the observation
that good conductors (copper, gold) tend not to become superconductors might
mean that superconductivity is produced by a relatively strong interaction between
the conduction electrons and the lattice vibrations, or phonons, in those metals that
were not good normal conductors. The Hamiltonian he produced to study the
question Ð the ÔÔFr\"{o}hlich HamiltonianÕÕ Ð implicitly contained the ÔÔisotope effect,ÕÕ
the prediction, later confirmed, that the superconducting transition temperature is
inversely proportional to the square root of the mass of the ions in the lattice
(Fr\"{o}hlich didnÕt realize he had made this prediction until he read about the
experiments that demonstrated the effect).$^{20}$ ''}

\end{document}